\documentstyle[12pt,epsf,aaspp4]{article}

\lefthead{Weinberg}
\righthead{Fundamental-plane ellipticals}

\def\msun{{\rm\,M_\odot}}
\def\kpc{{\rm\,kpc}}

\newdimen\hwidth
\newdimen\colspace
\hwidth=\textwidth \divide\hwidth by 2 \advance\hwidth-2\colspace

\begin{document}

\title{The fate of cannibalized fundamental-plane ellipticals}

\author{Martin D. Weinberg\altaffilmark{1}}

\affil{Department of Physics and Astronomy\\ University of
Massachusetts, Amherst, MA 01003-4525\\ weinberg@phast.umass.edu}

\altaffiltext{1}{Alfred P. Sloan Foundation Fellow.}

\begin{abstract}
Evolution and disruption of galaxies orbiting in the gravitational
field of a larger cluster galaxy are driven by three coupled
mechanisms: 1) the heating due to its time dependent motion in the
primary; 2) mass loss due to the tidal strain field; and 3) orbital
decay.  Previous work demonstrated that tidal heating is effective
well inside the impulse approximation limit.  Not only does the
overall energy increase over previous predictions, but the work is
done deep inside the secondary galaxy, e.g. at or inside the half mass
radius in most cases.  Here, these ideas applied to cannibalization of
elliptical galaxies with fundamental-plane parameters.

In summary, satellites which can fall to the center of a cluster giant
by dynamical friction are evaporated by internal heating by the time
they reach the center.  This suggests that true merger-produced
multiple nuclei giants should be rare.  Specifically, secondaries with
mass ratios as small as 1\% on any initial orbit evaporate and those
on eccentric orbits with mass ratios as small as 0.1\% evolve
significantly and nearly evaporate in a galaxian age.  Captured
satellites with mass ratios smaller than roughly 1\% have insufficient
time to decay to the center.  After many accretion events, the model
predicts that the merged system has a profile similar to that of the
original primary with a weak increase in concentration.
\end{abstract}


\section{Introduction} \label{sec:intro}

The current picture galaxy evolution in clusters leads naturally to
galactic cannibalism, especially deep in the potential well where the
giants reside.  Although multiple nuclei candidates have been
identified (e.g. Tonry 1985, Lauer 1988\nocite{Tonr:85,Laue:88}),
recent searches turned up many fewer inner core objects than expected
(Tremaine 1995\nocite{Trem:95}).

A closer look at the evolutionary picture is motivated by the recent
demonstration that heating of galaxies or star clusters due the
time-dependent tidal field can drive their evolution at a rate beyond
impulse approximation estimates (Weinberg 1994abc, hereafter W1--3
\nocite{Wein:94a,Wein:94b,Wein:94c}, Murali \& Weinberg 1996ab,
hereafter MW1--2\nocite{MuWe:96a,MuWe:96b}).  This theory invalidates
the following often-used argument. The higher density of satellite
galaxies implies shorter internal orbital times than the orbit of the
satellite itself.  Therefore, the stellar orbits in the satellite will
adiabatically invariant to the tidal force, and since the dynamical
friction time scale is much less than a Hubble time for galaxy masses
above $10^9 \msun$, the satellite should sink to the center without
suffering tidal disruption and remain a distinct compact entity.  This
paper presents estimates of the evolutionary path and evaporation
lifetime for the cannibalized fundamental-plane elliptical galaxies.
We will find that ellipticals with sufficient mass to decay are heated
and evaporated before a multiple nucleus system can result, although
such systems may exist transiently.  The results also illustrate the
interplay between tidal heating, tidal stripping and orbital decay.
The likely evaporation of accreted galaxies may help reconcile the
observation of a bimodal velocity distribution of multiple nuclei
(Tonry 1985) as 1) a dynamical friction mediated selection effect
(Merritt 1984) or recently captured secondary and 2) a transient
population of evaporating secondaries.

We begin with a description of the astronomical scenario in
\S\ref{sec:model}.  All members of the fundamental plane are
represented by a spherical model with fixed concentration; this is
consistent with the observed fundamental plane relations given the
Faber-Jackson (1976) relation although the best estimates suggest a
weak dependence on concentration.  We assume that the secondary is
captured from the cluster by dynamical drag and consider evolution
after the secondary is bound to the primary.  The important dynamical
ingredients and their implementation is briefly discussed in
\S\ref{sec:method}; the technical details can be found in the Appendix
and elsewhere (W2, MW2).  The results for a number astronomical
scenarios are described in \S\ref{sec:results}; these include the
survival and evolution as a function of mass, orbital decay, and the
resulting distribution of stripped stars in the primary.  A summary
and discussion is presented in \S\ref{sec:summary}.

\section{Astronomical scenario} \label{sec:model}

\subsection{Background profile and fundamental plane scaling} \label{sec:scale}

I have chosen a King model for both the primary and secondary.  King
models with $\log c=2.35$ are representative elliptical profiles
(e.g. Mihalas \& Binney 1981, Vader \& Chaboyer 1994\nocite{VaCh:94})
although King models are not good fits in all cases\footnote{The
concentration parameter is defined as $\log
c\equiv\log_{10}(R_{max}/R_{core})$.}.  Nonetheless, the mass model
parameterizes the range of stellar orbital times, and this range
determines the overall evolution rate from the resonant heating
process to be described below.  An appropriate concentration ensures
that a realistic range of orbital time scales are included.  The
conclusions (\S\ref{sec:results}) are weakly dependent on the inner
profile and other fine details of the model.

The radius and mass concentration chosen according to two fundamental
plane relations.  The first is based on the virial theorem and the
Faber-Jackson relation, $L\propto\sigma^4$ (Faber \& Jackson
1976\nocite{FaJa:76}), which results in the following scaling:
\begin{equation}
	R \propto M^{ 1/2}.
\end{equation}
and $\rho\propto M^{-1/2}$.  The concentration parameter is invariant
under any fundamental plane scaling also assuming the Faber-Jackson
relation.  Therefore all three King model parameters, mass, tidal or
maximum radius, and concentration, are fixed for each secondary of
given mass.  The second is based on recent observed fundamental plane
relations (e.g. Pahre et al. 1995\nocite{PaDdC:95}, Faber
1995\nocite{Fabe:95}):
\begin{equation}
	R \propto M^{0.9}.
\end{equation}
To reduce the overall number of parameters in this study, I have
chosen to retain the Faber-Jackson relation and constant-concentration
models, even though recent fundamental relations predict that central
density and therefore concentration class scales with mass.  Changes
in concentration predominantly change the inner profile and, as noted,
only weakly effect the overall evolutionary track of the accreted
secondary.

\begin{figure}
\begin{center}
\mbox{\epsfxsize=\textwidth\epsfbox{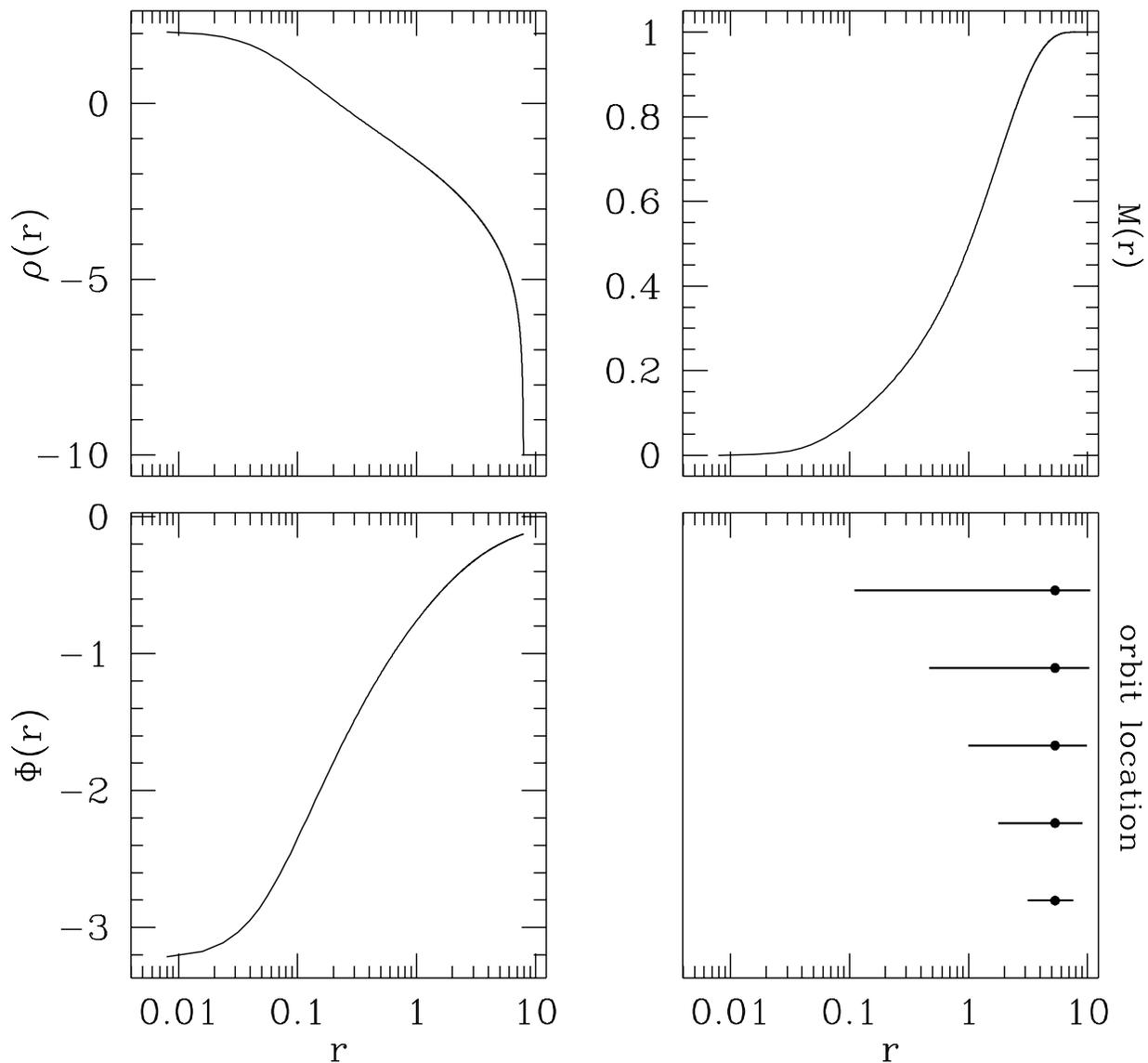}}
\caption{\label{fig:profile} Density, mass and potential for $W_0=9.5$
King model in dimensionless units.  The model is nearly isothermal for
$0.05\lesssim r\lesssim5$.  The diagram at the lower right shows the
pericenter and apocenter radii for each orbit (ends of segments) and
guiding center (circular orbit) radii (solid dots).}
\end{center}
\end{figure}

There are three remaining parameters: orbital energy, orbital
eccentricity and secondary to primary mass ratio.  Orbital evolution
is determined using local dynamical friction which requires the
secondary to be inside the primary (see \S\ref{sec:chandra}).  The
initial orbits for the secondaries, then, are chosen to have an energy
whose circular orbit encloses the 99\% of the primary mass.  In other
words, we consider evolution just subsequent to capture. Eccentricity
is parameterized by the ratio of orbital angular momentum to the
maximum defined by the energy of the orbit, $\kappa\equiv
J/J_{max}(E)$ and five values are chosen: $0.1(0.2)0.9$.  A pure
circular (radial) orbit has $\kappa=1.0$ ($\kappa=0.0$).  Because a
captured elliptical is likely to be on a eccentric orbit, we will
emphasize the $\kappa=0.1$ case.  The model profile and location of
initial orbits in the model is shown in Figure \ref{fig:profile}.
Finally, each set of 5 orbits is evolved for 4 different secondary to
primary mass ratios: $10^{-4}$, $10^{-3}$, $10^{-2}$, and $10^{-1}$.

\begin{figure}[thb]
\begin{center}
\mbox{\epsfxsize=3.0in\epsfbox{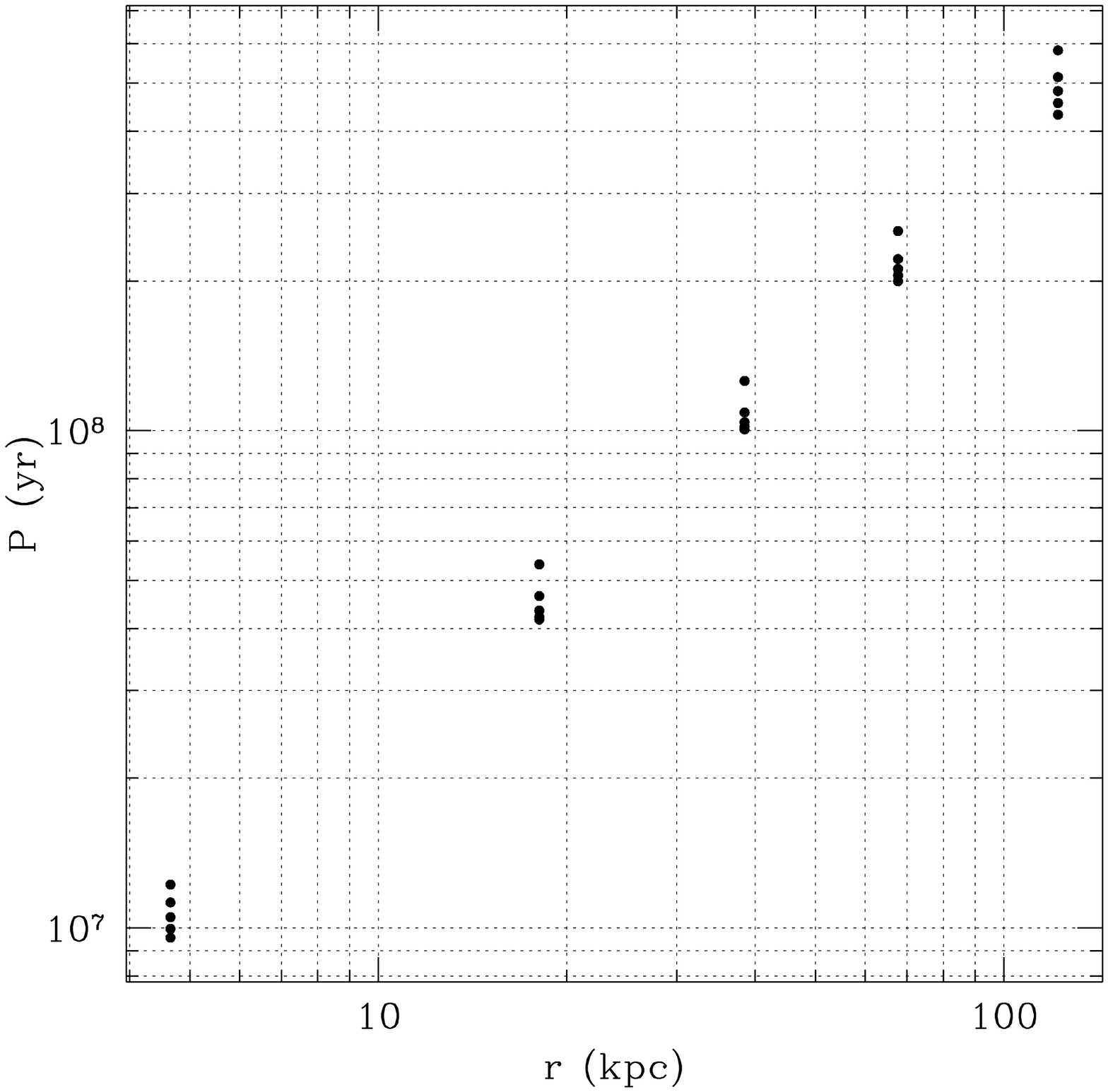}}
\caption{\label{fig:period} Periods of orbit scaled to a central
cluster galaxy with $M=10^{14}\msun$ and $R_{max}=300\kpc$.  Dots
represent the orbital periods with the guiding center radius enclosing
10(20)90\% of the primary mass.  The values $\kappa=0.1(0.2)0.9$ are
ordered from bottom to top.}
\end{center}
\end{figure}

Dimensionless units are chosen for the King model such that $G=M=1$
and total gravitational potential energy $W=-1/2$.  For the $W_0=9.5$
King model, $R_{core}=0.5$ with outer radius $R_{max}=7.91$ in these
units.  I will take a fiducial central cluster galaxy to have
$M=10^{14}\msun$ inside of $R_{max}=300\kpc$ and which sets the time
scale quoted in years in \S\ref{sec:results}.  This fiducial choice is
similar to that for M87 (e.g. Binney \& Tremaine 1987, Merritt \&
Tremblay 1993\nocite{BiTr:87,MeTr:93}).  A different choice simply
shifts the quoted time scale by the ratio
\begin{equation} \label{eq:tscale}
	{T\over T_o} = \left({M\over 10^{14}\msun}\right)^{-1/2}
	\left({R_{max}\over 300 \kpc}\right)^{3/2}.
\end{equation}
For reference, orbital periods for the fiducial scaling whose guiding
center radii enclose 10(20)90\% for $\kappa=0.1(0.2)0.9$ are described
in Figure \ref{fig:period}.

\section{Method overview} \label{sec:method}

Evolution in the cannibalized ellipticals is caused by the following
four interacting physical effects: 1) resonant heating and orbit
shocking; 2) self-consistent gravity; 3) tidal stripping; and 4)
dynamical friction.  These will be briefly described below and in the
Appendix.  We will see that dependencies in the effect of the four
physical processes govern the subsequent evolution.

\subsection{Resonant heating} \label{sec:heating}

The orbiting secondary galaxy experiences a differential or {\it
tidal}\/ force.  The combined strain and compressive force is
time-dependent and can do work on the secondary galaxy.  If the change
in tidal force is rapid compared to internal orbital time scales, a
{\it gravitational shock}, the work can be computed using the impulse
approximation.  However even if the change in strain is slower than
internal orbital time scales, significant work may still be done: most
realistic galaxies will have resonances between the two (or more)
internal orbital frequencies the external forcing frequency which
leads to significant energy and angular momentum exchange (W1--3).

More picturesquely, the time-dependent force will excite a wake in the
secondary.  The wake will be dominated by a quadrupole or bar-like
distortion whose pattern speed is determined by the external
frequency.  Similar to torquing by spiral arms, this `bar' then
couples to the tidal force, transferring energy and angular momentum
to resonant orbits.  The perturbation-theory-derived heating rates
used here are in good agreement with n-body simulations (cf. MW2,
Johnston et al. 1996\nocite{JoHW:96}) with the advantage of being able
to follow a weak disturbance without noise.

\subsection{Self-consistent gravity}

By Jeans' theorem (e.g. Binney \& Tremaine 1987\nocite{BiTr:87}), an
equilibrium of regular orbits is described by a phase-space
distribution function, $f=f({\bf I})$ where ${\bf I}$ are the actions
(or energy and angular momentum for a spherical system).  The
associated potential and density solve the Poisson equation by
construction.  Although the actions of most orbits are invariant to
the slowly changing tidal strain, the resonant heating described above
changes the actions of some small subset of orbits resulting in a
slightly out of equilibrium system.

At regular intervals, the Poisson equation is iteratively solved to
maintain equilibrium.  Because the external force is assumed to do
negligible work on an internal orbital time scale, all external
perturbations may be temporarily turned off which fixes the actions
and simplifies solution.  So the overal evolution consists of two
phases: 1) evolution of phase space due to external perturbations in a
fixed gravitational potential; and 2) dynamical readjustment with all
perturbations removed.  Practically, a new equilibrium is only
computed when the changing phase-space distribution implies a 1--2\%
change to the stellar orbits.

As the equilibrium profile evolves, new orbits become resonant with
the external force.  In this way, a small set of resonant orbits at
any one time can change the global structure over a number of
dynamical time scales.  Finally for the results below, the resulting
equilibrium phase-space distribution is forced to be spherical and
isotropic.  This is not an in-principle demand---the numerical
implementation is general---but a choice driven by available CPU time.

\subsection{Tidal stripping}

The outer boundary of a secondary is defined by the points at which a
star is more strongly attracted by the primary.  For a circular orbit,
this point is the analogous inner Lagrange point in the restricted
three-body problem.  However, for an eccentric orbit, this is not an
easily parameterized problem; these points change as the secondary
orbits resulting in foliated stable and unstable regions (e.g. Keenan
1981\nocite{Keen:81}).  N-body simulations suggest that setting the
boundary to the inner Lagrange point at perigalacticon is a fair
prescription.

The location of the inner Lagrange point scales with the ratio of mean
density of the secondary to mean density of the primary enclosed with
the secondary's orbit.  Therefore, as the secondary evolves due to
time-dependent heating as described in \S\ref{sec:heating}, stars may
find themselves on the unbound side of the tidal limit.  This loss of
material also changes the equilibrium.  If too much material is
evaporated, global equilibrium may be lost and the smaller galaxy
``disrupts''.

\subsection{Dynamical friction} \label{sec:chandra}

Finally, the orbit itself is evolving by dynamical friction.  For
small secondaries, Chandrasekhar's dynamical friction formula is an
acceptable approximation (Chandrasekhar 1943\nocite{Chan:43a}, see
e.g. Binney and Tremaine 1987\nocite{BiTr:87}).  This approximation
assumes that the primary is infinite and homogeneous with the local
value of density and distribution of velocities.  The drag force is
anti-parallel to the motion of the secondary assuming velocity
isotropy.  For large secondaries, the situation is more complex
(e.g. Hernquist \& Weinberg 1989\nocite{HeWe:89}, Weinberg
1989\nocite{Wein:89}) but the local approximation will be used for
simplicity.  Because the evolution of large secondaries is rapid, it
is unlikely that this assumption affects any conclusion.  Further
consequences of the decaying orbit are an increasing resonant heating
rate and stronger tidal limit, both of which accelerate the evolution.

\section{Evolution of satellite galaxies} \label{sec:results}

The models and methods of \S\ref{sec:model} and \S\ref{sec:method} are
applied to groups of twenty models each.  Each group of twenty has
four mass ratios, $M_{ratio}=10^{-1}, 10^{-2}, 10^{-3}, 10^{-4}$ and
five eccentricities, $\kappa=0.1(0.2)0.9$
(cf. Fig. \ref{fig:profile}).  The two groups discussed here have
guiding center orbits which enclose 99\% of the primary mass.  The
first group uses the virial scaling and the second uses the observed
fundamental plane scaling (cf.  \S\ref{sec:model}).

Recall that the physical times quoted below assume a primary mass of
$10^{14}\msun$ inside of $300\kpc$.  Equation (\ref{eq:tscale}) may be
used to scale to any desired primary mass and radius.

\subsection{Disruption and Survival}

\begin{figure}[thb]
\begin{center}
\mbox{
\mbox{\epsfxsize=\hwidth\epsfbox[75 75 385 277]{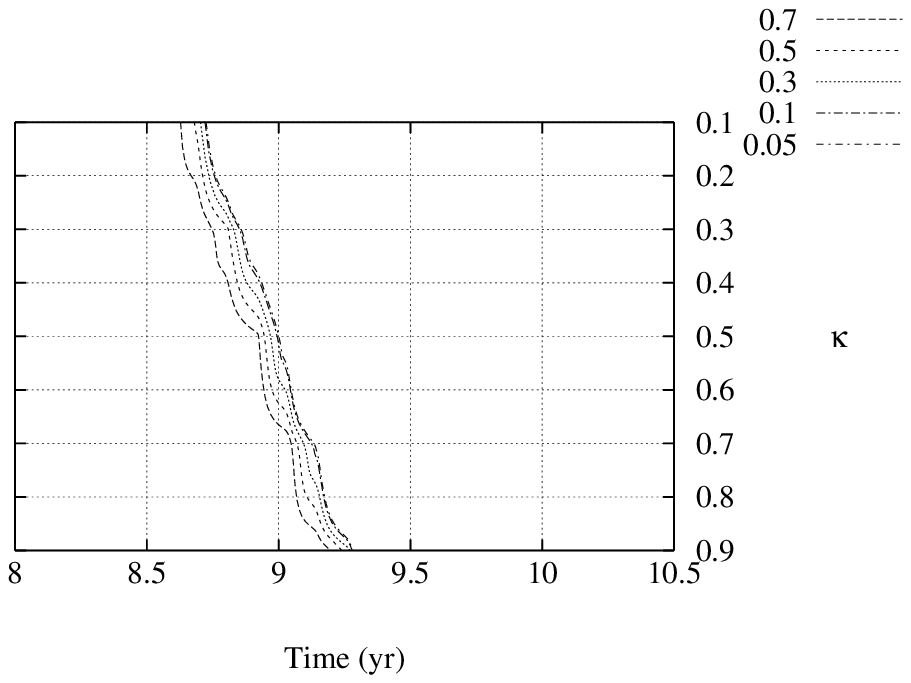}}
\mbox{\epsfxsize=\hwidth\epsfbox[75 75 385 277]{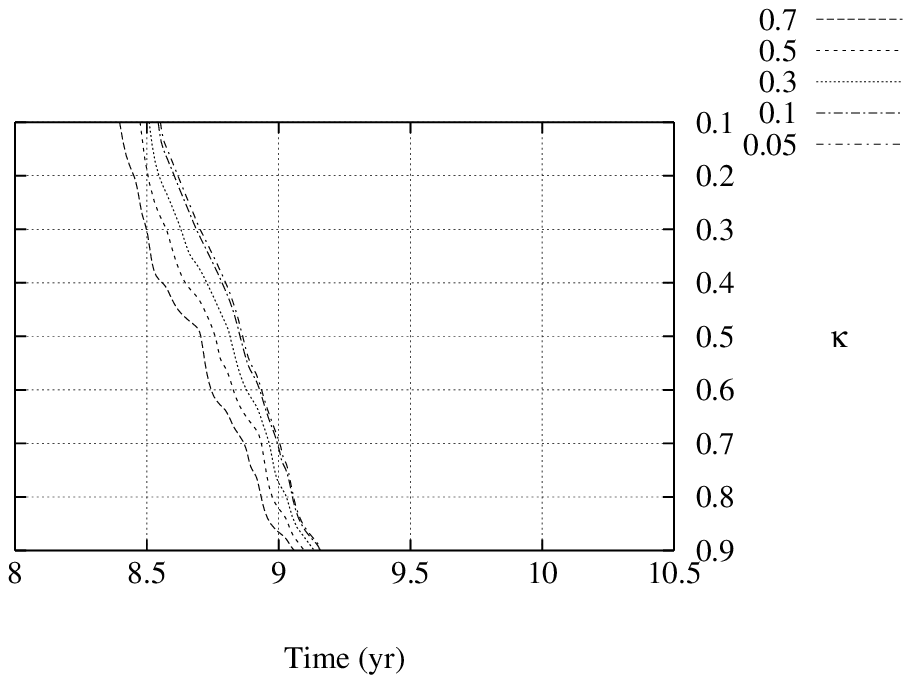}}
}
\caption{\label{fig:mass1} Contours show remaining mass fraction (key
at upper left) as a function of time (logarithmic scale) and initial
value of $\kappa=J/J_{max}$.  The secondary to primary mass ratio is
$10^{-1}$.  The right hand (left hand) plot shows the virial
(observed) fundamental plane scaling.  The scalloping in the contours
is caused by the projection of a finite grid.}
\end{center}
\end{figure}

\begin{figure}[p]
\begin{center}
\mbox{
\mbox{\epsfxsize=\hwidth\epsfbox[75 75 385 277]{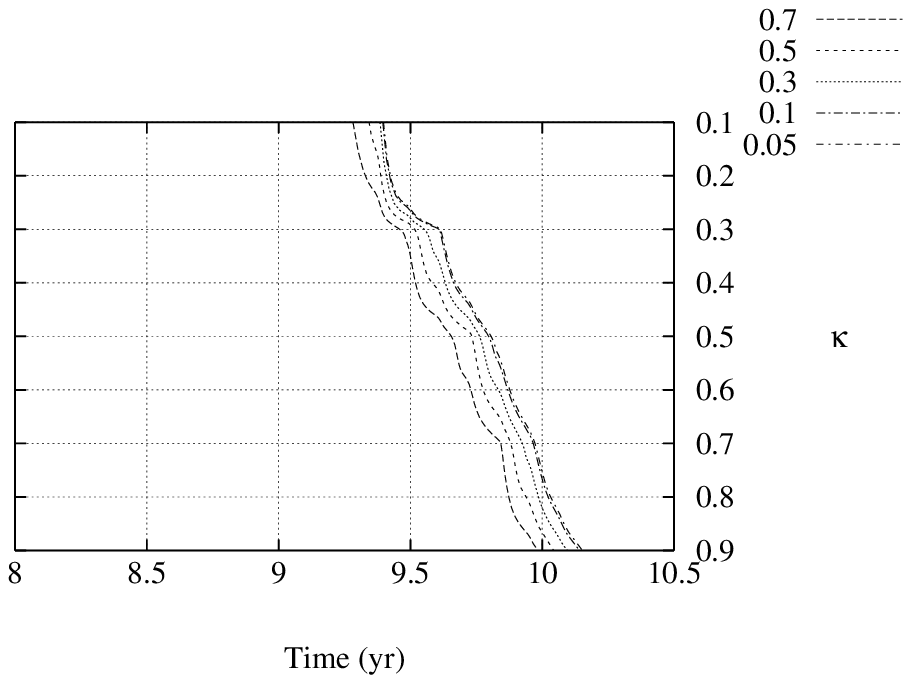}}
\mbox{\epsfxsize=\hwidth\epsfbox[75 75 385 277]{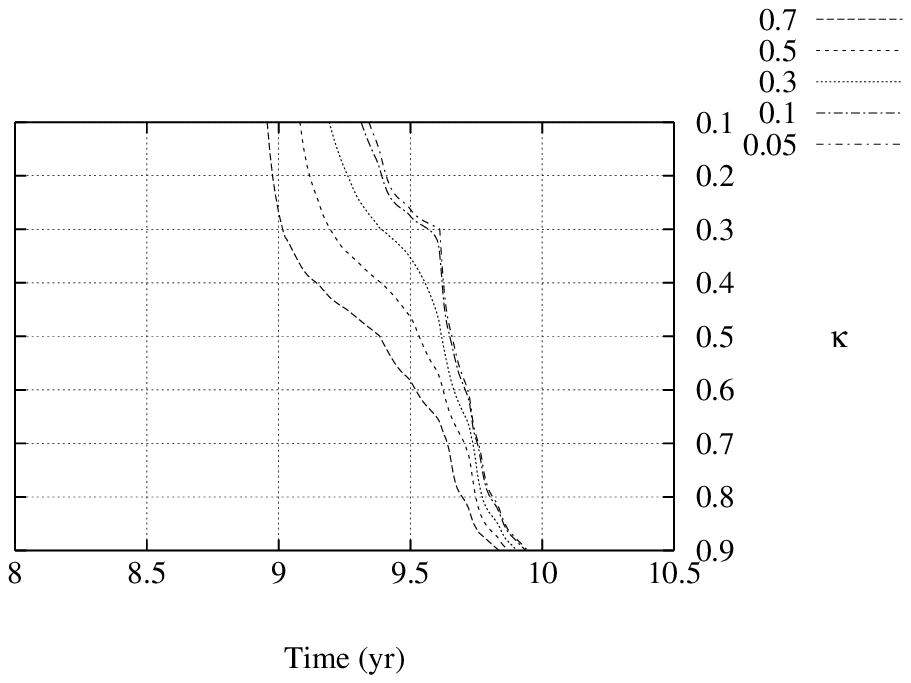}}
}
\caption{\label{fig:mass2} As in Fig. \protect{\ref{fig:mass1}} but for mass
ratio $10^{-2}$.}
\mbox{
\mbox{\epsfxsize=\hwidth\epsfbox[75 75 385 277]{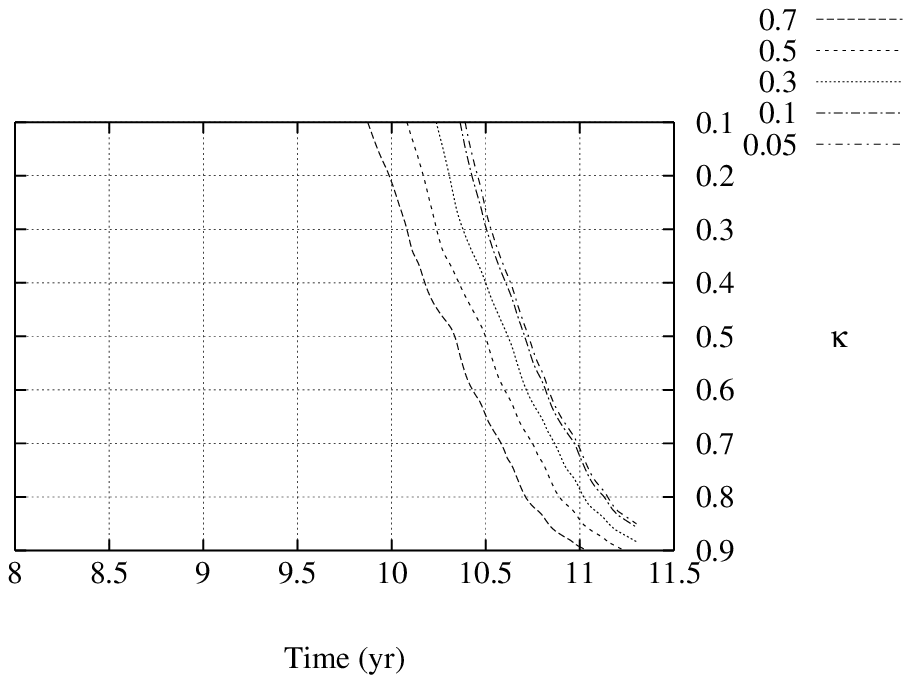}}
\mbox{\epsfxsize=\hwidth\epsfbox[75 75 385 277]{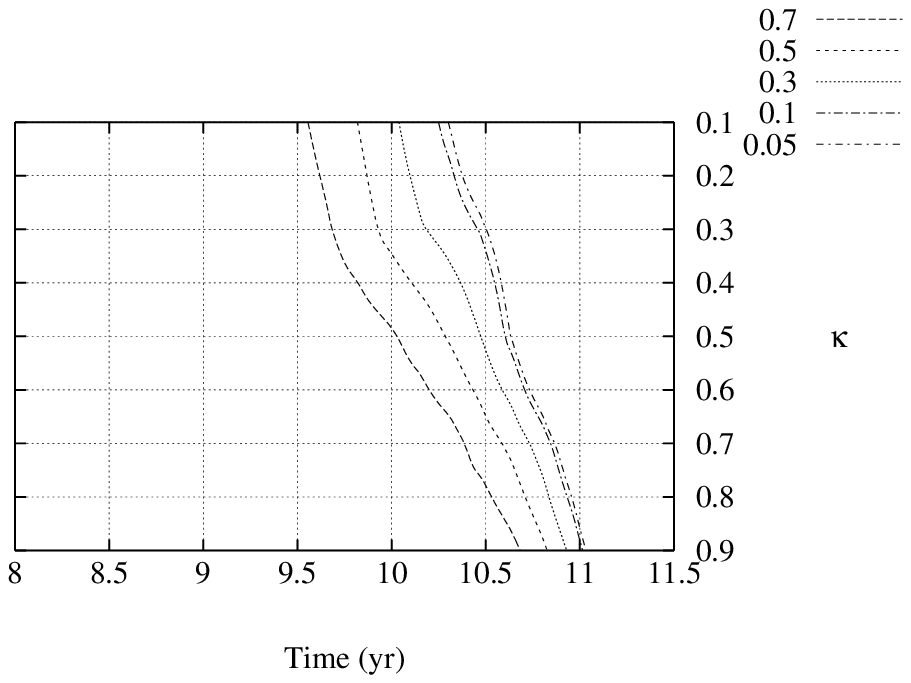}}
}
\caption{\label{fig:mass3} As in Fig. \protect{\ref{fig:mass1}} but for mass
ratio $10^{-3}$.  Range in time is extended to accommodate alternative
scalings.}
\mbox{
\mbox{\epsfxsize=\hwidth\epsfbox[75 75 385 277]{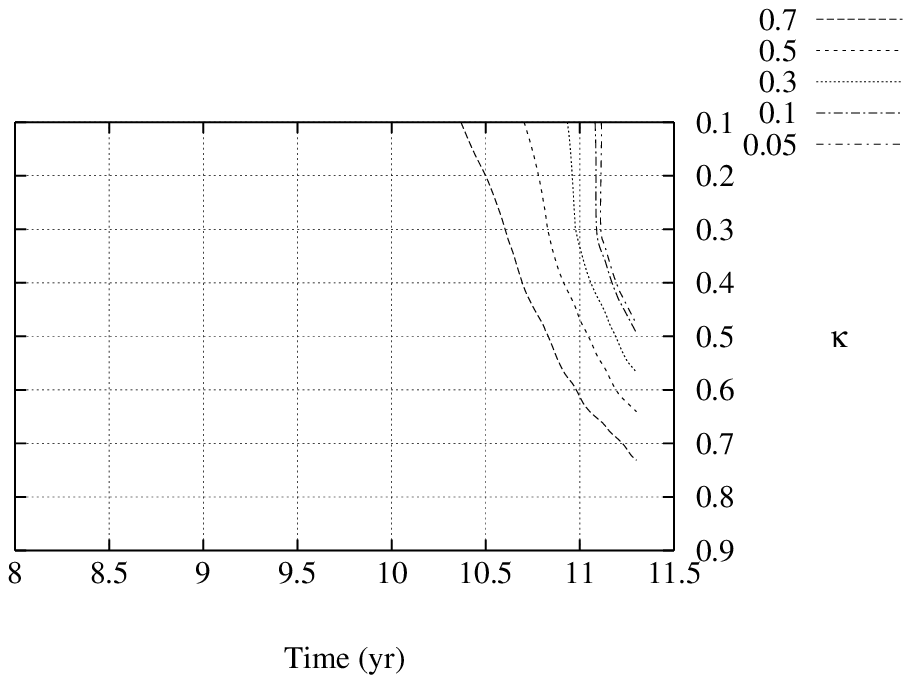}}
\mbox{\epsfxsize=\hwidth\epsfbox[75 75 385 277]{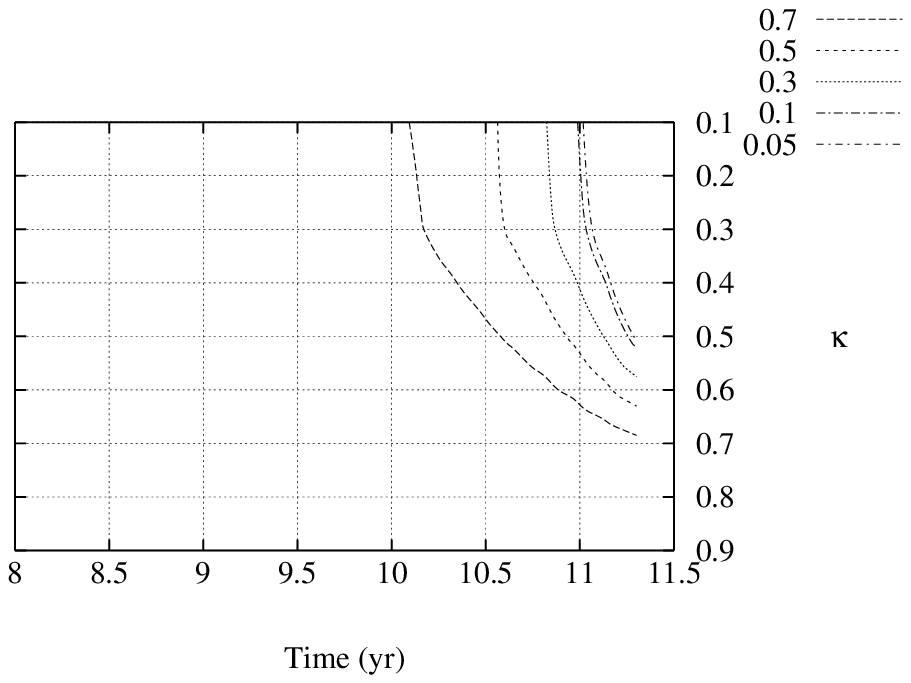}}
}
\caption{\label{fig:mass4} As in Fig. \protect{\ref{fig:mass1}} but for mass
ratio $10^{-4}$.  Range in time is extended to accommodate alternative
scalings.}
\end{center}
\end{figure}

Figures \ref{fig:mass1}--\ref{fig:mass4} describe the mass evolution
for fundamental plane ellipticals as a function of initial
eccentricity for the four mass ratios.  The contours indicate the mass
fraction remaining at the time indicated.  The vertical axis shows
increasing initial orbital eccentricity.  Galaxies to the right of the
0.05 contour have completely evaporated.  Heating and stripping is
severe for the most eccentric orbits and the highest mass ratios.  For
the ratio $10^{-1}$, the orbit decays quickly and has evaporated for
all eccentricities by roughly 1 Gyr (Fig. \ref{fig:mass1}).  We will
see in \S{\ref{sec:orbdecay}} that disruption occurs near the center
of the host galaxy.

The trends are similar for smaller mass ratios.  A 1\% secondary
(Fig. \ref{fig:mass2}) evaporates in 10 Gyr for a nearly circular
orbits and in roughly 2 Gyr for an eccentric orbit.  A 0.1\%
secondary---a large dwarf galaxy---does not completely evaporate in 10
Gyr even for an eccentric orbit, although it is close.  Evolution is
slower for smaller mass ratios because 1) the density of the secondary
is larger and therefore couples more weakly to the tidal field; and 2)
the orbital decay rate, which is proportional to the mass ratio, is
slower.

\subsection{Orbital decay} \label{sec:orbdecay}

\begin{figure}[p]
\begin{center}
\mbox{
\vbox{
	\hbox{
		\hbox{\epsfxsize=\hwidth\epsfbox{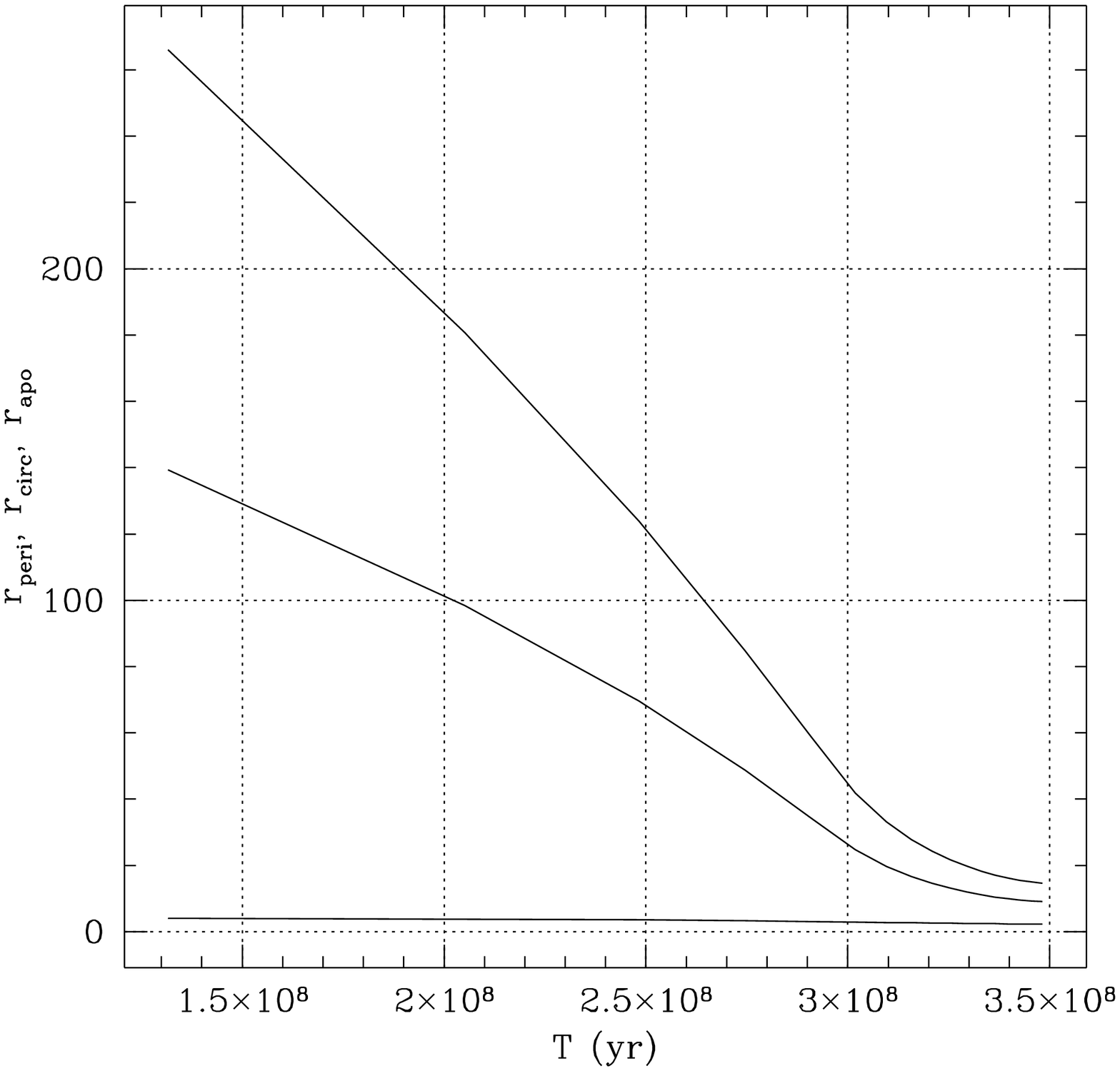}}
		\hbox{\epsfxsize=\hwidth\epsfbox{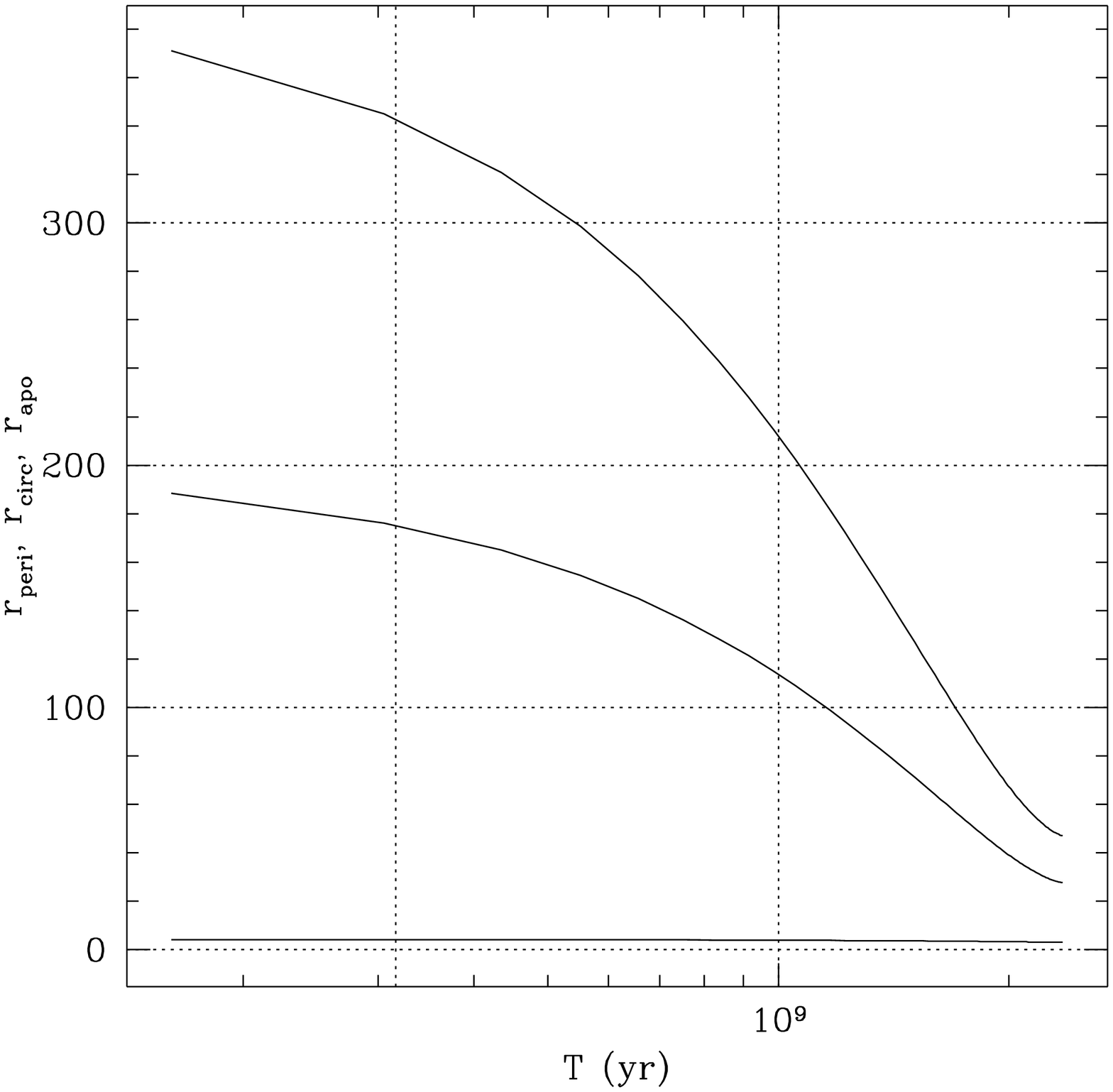}}
	}
	\hbox{
		\hbox{\epsfxsize=\hwidth\epsfbox{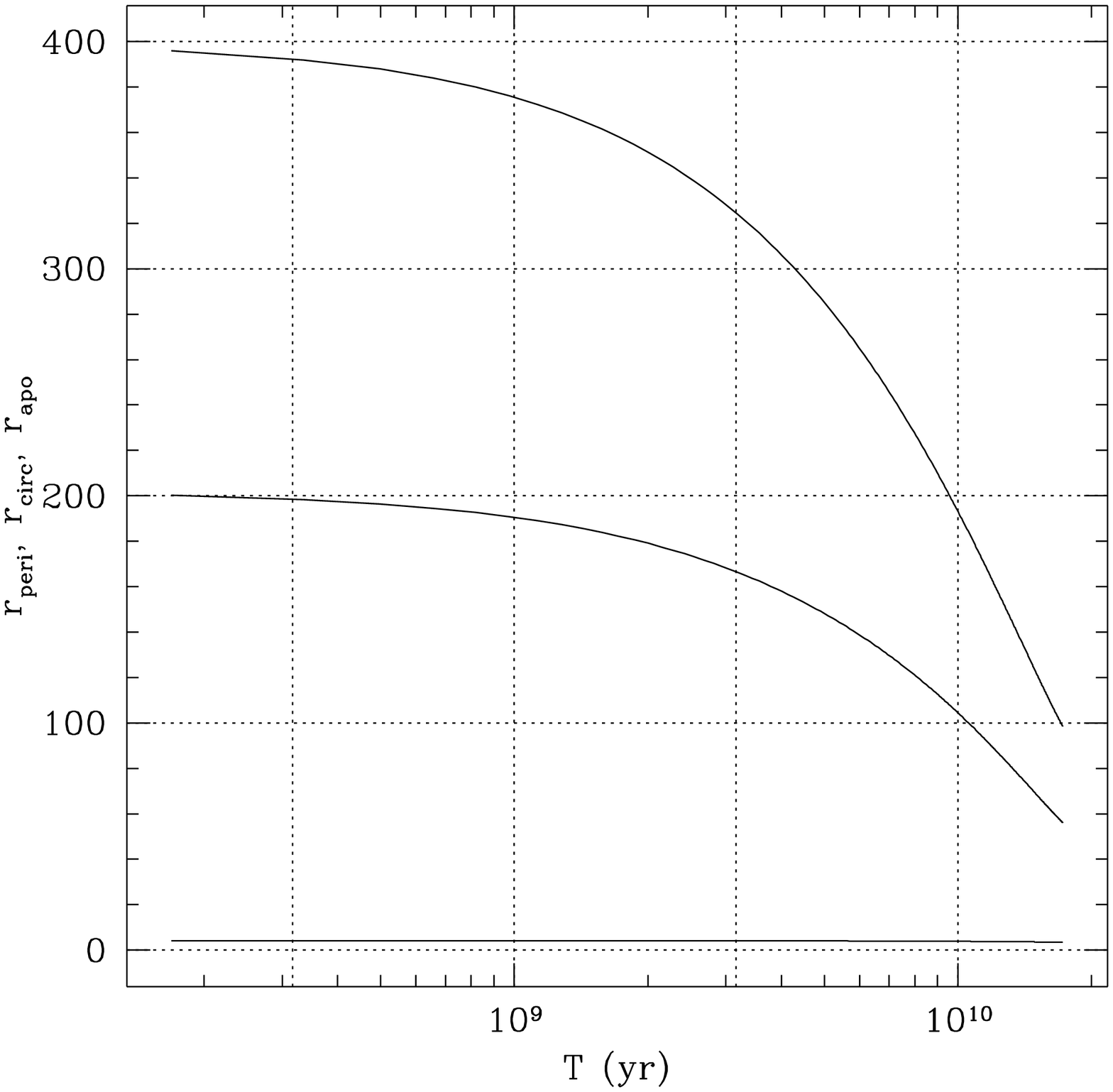}}
		\hbox{\epsfxsize=\hwidth\epsfbox{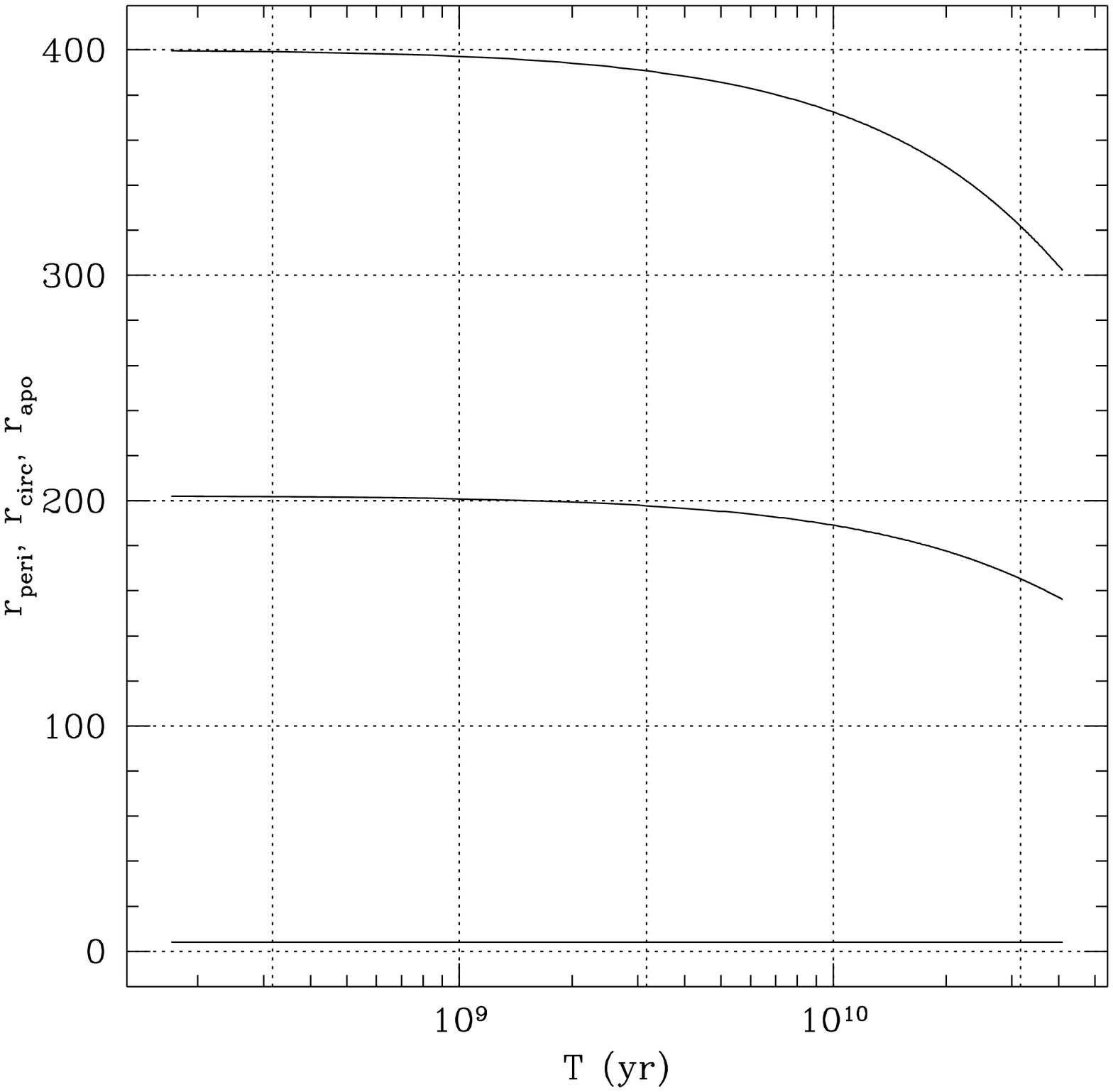}}
	}
}}
\caption{\label{fig:orbdecay} Orbital evolution for initial
$\kappa=0.1$ orbits for the four mass ratios $10^{-1}, 10^{-2},
10^{-3}, 10^{-4}$ (right to left, top to bottom) with observed
fundamental plane scaling (cf. Figs.
\protect{\ref{fig:mass1}}---\protect{\ref{fig:mass4}}).  Satellites
with $M_{ratio}=10^{-1}$ and $10^{-2}$ have zero mass at the final
time shown.}
\end{center}
\end{figure}

\begin{figure}[thb]
\begin{center}
\mbox{\epsfxsize=3.0in\epsfbox{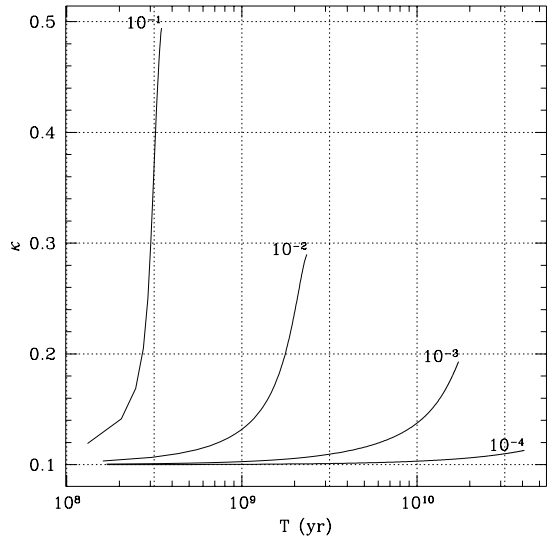}}
\caption{\label{fig:circularize} Change in $\kappa=J/J_{max}$ for mass
ratios indicated for orbits with $\kappa=0.1$ initially.}
\end{center}
\end{figure}

\begin{figure}[p]
\begin{center}
\mbox{
\vbox{
	\hbox{
		\hbox{\epsfxsize=\hwidth\epsfbox{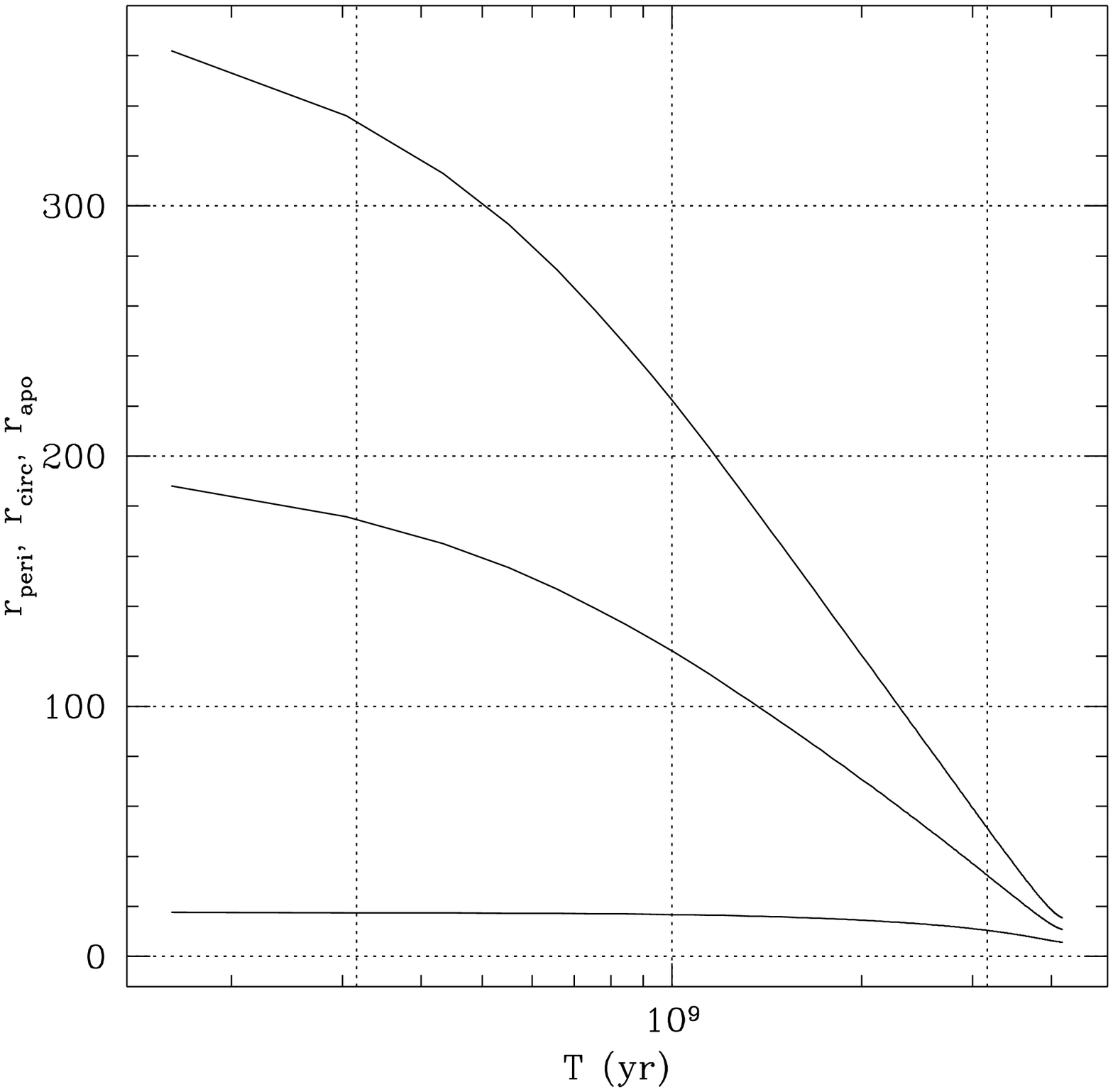}}
		\hbox{\epsfxsize=\hwidth\epsfbox{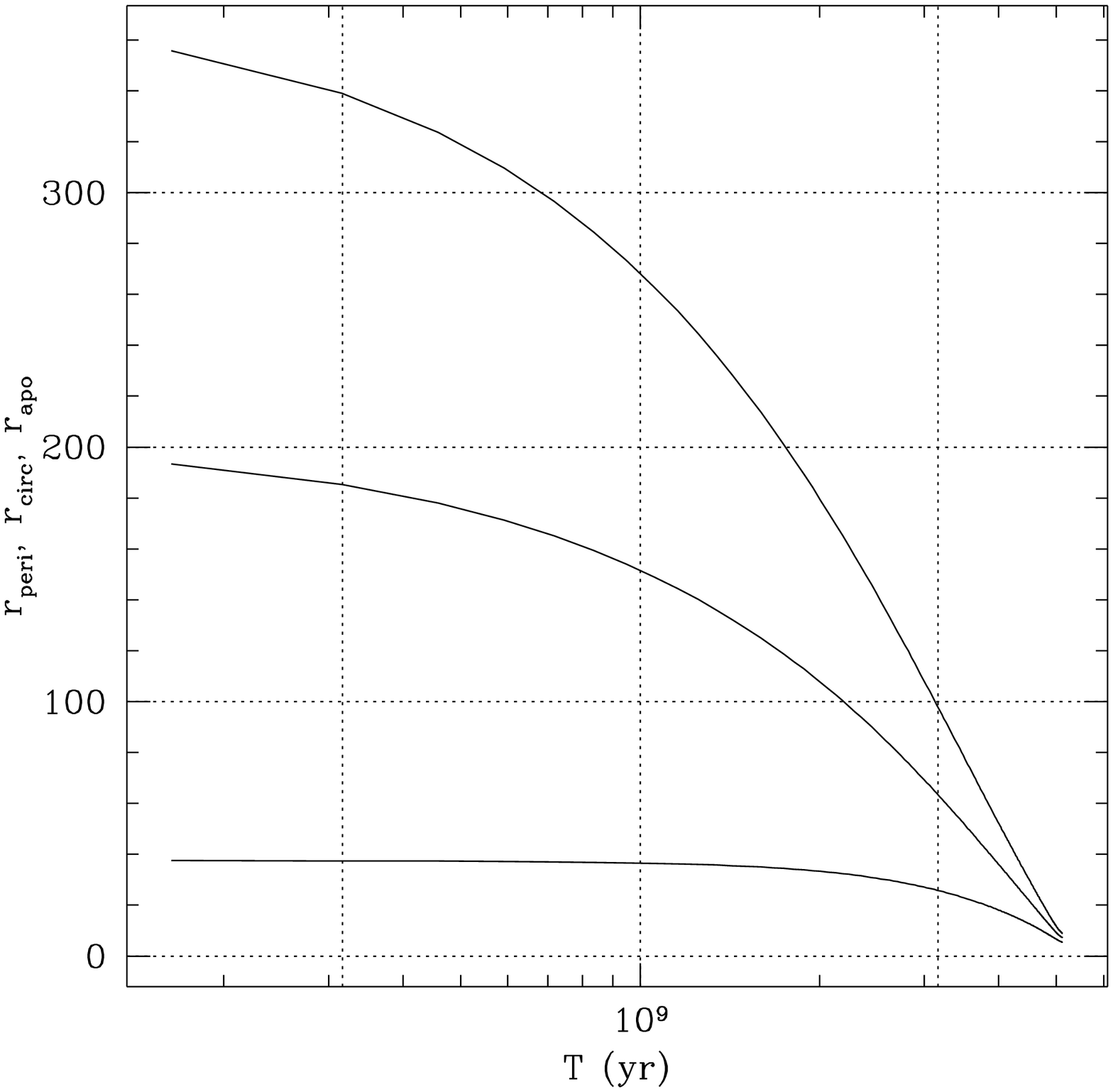}}
	}
	\hbox{
		\hbox{\epsfxsize=\hwidth\epsfbox{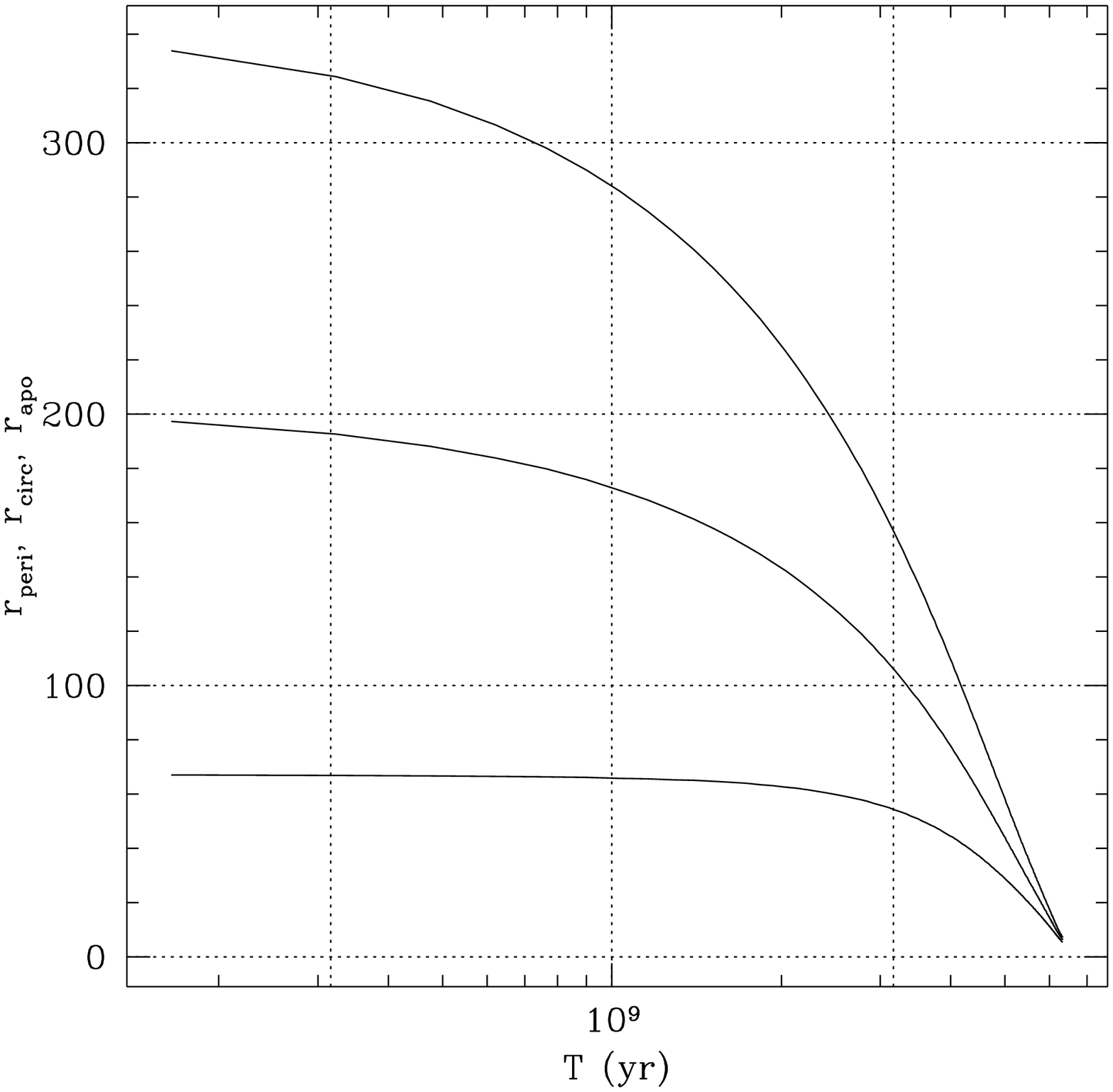}}
		\hbox{\epsfxsize=\hwidth\epsfbox{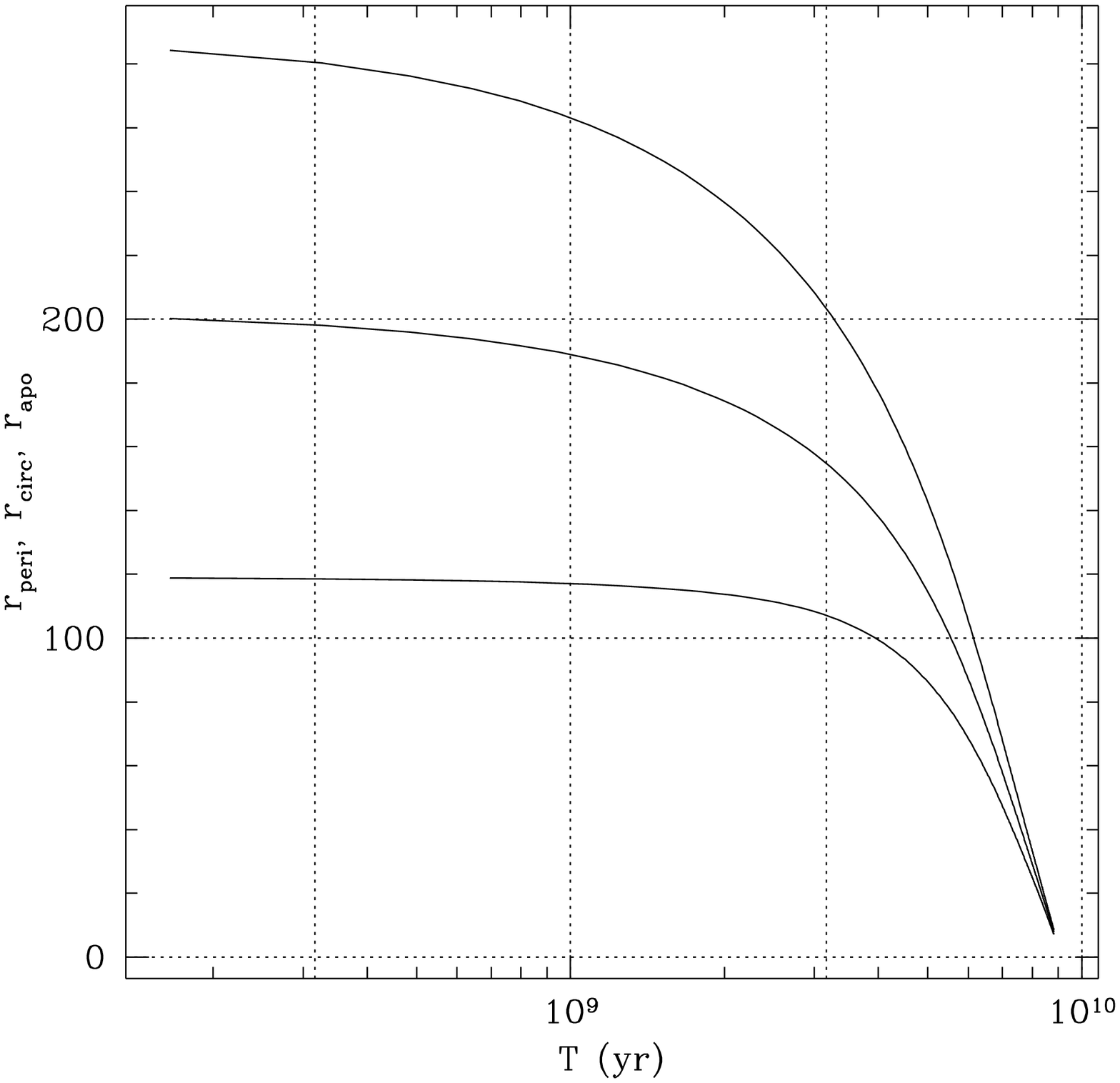}}
	}
}}
\caption{\label{fig:orbdkap} Orbital evolution for initial
$\kappa=0.3, 0.5, 0.7, 0.9$ orbits for mass ratios $10^{-2}$,
with observed fundamental plane scaling.}
\end{center}
\end{figure}

Figure \ref{fig:orbdecay} shows the orbital evolution of initially
$\kappa=0.1$ orbits for the four mass ratios.  The decay rate is
computed using Chandrasekhar's formula with
$\ln\Lambda=\ln(R_{max}/r_{1/2})$ where $r_{1/2}$ is the current
half-mass radius of the secondary. Orbital torques are also computed
in the local approximation with Chandrasekhar's formula.  These
eccentric orbits becomes more circular during its decay
(Fig. \ref{fig:circularize}) as previously described by Bontekoe \&
van Albada (1987\nocite{BovAl:87}).  Initially, the $\kappa=0.1$
orbits with guiding center trajectories enclosing 99\% of the primary
mass have apocenters outside the primary (cf. Fig. \ref{fig:profile})
which may lead to an overestimate of the decay time.  For the fiducial
model ($M=10^{14}\msun, R_{max}=300\kpc$), only the 10\% and 1\% mass
ratio secondaries can decay into the center in roughly 10 Gyr.  The
decay time for lower mass galaxies is increased by the concurrent mass
loss.

The longer lifetimes for large initial $\kappa$ is correlated with the
longer orbital decay times.  This is shown in Figure \ref{fig:orbdkap}
which describes the evolution of 1\% secondaries for
$\kappa=0.3(0.2)0.9$ (the second panel in Fig. \ref{fig:orbdecay} is
the first member of this sequence with $\kappa=0.1$).  The decay rate
is nearly constant for orbits in the inner primary (roughly inside the
half-mass radius of $60\kpc$).  The rate increases with decreasing
initial eccentricity because the secondary spends a larger fraction of
its time at higher primary density.  This trend decreases the spread
of decay times with eccentricity, but does not compensate for the
slower initial evolution of low-eccentricity orbits.

Similarly, the steep gradient in time across the mass contours in
Figures \ref{fig:mass1}--\ref{fig:mass4} reflects the rapid mass
evolution which takes place during the final stage of orbital decay.
This trend is maintained at the smallest mass ratios although full
decay takes longer than a galaxian age for the fiducial scaling.

Combining the results of this and the previous subsection, we reach
the conclusion that {\it satellites which can fall to the center of a
cluster giant by dynamical friction are evaporated by internal
heating.}

\subsection{Distribution of stripped material}

\begin{figure}[p]
\begin{center}
\mbox{
\vbox{
	\hbox{
		\hbox{\epsfxsize=\hwidth\epsfbox{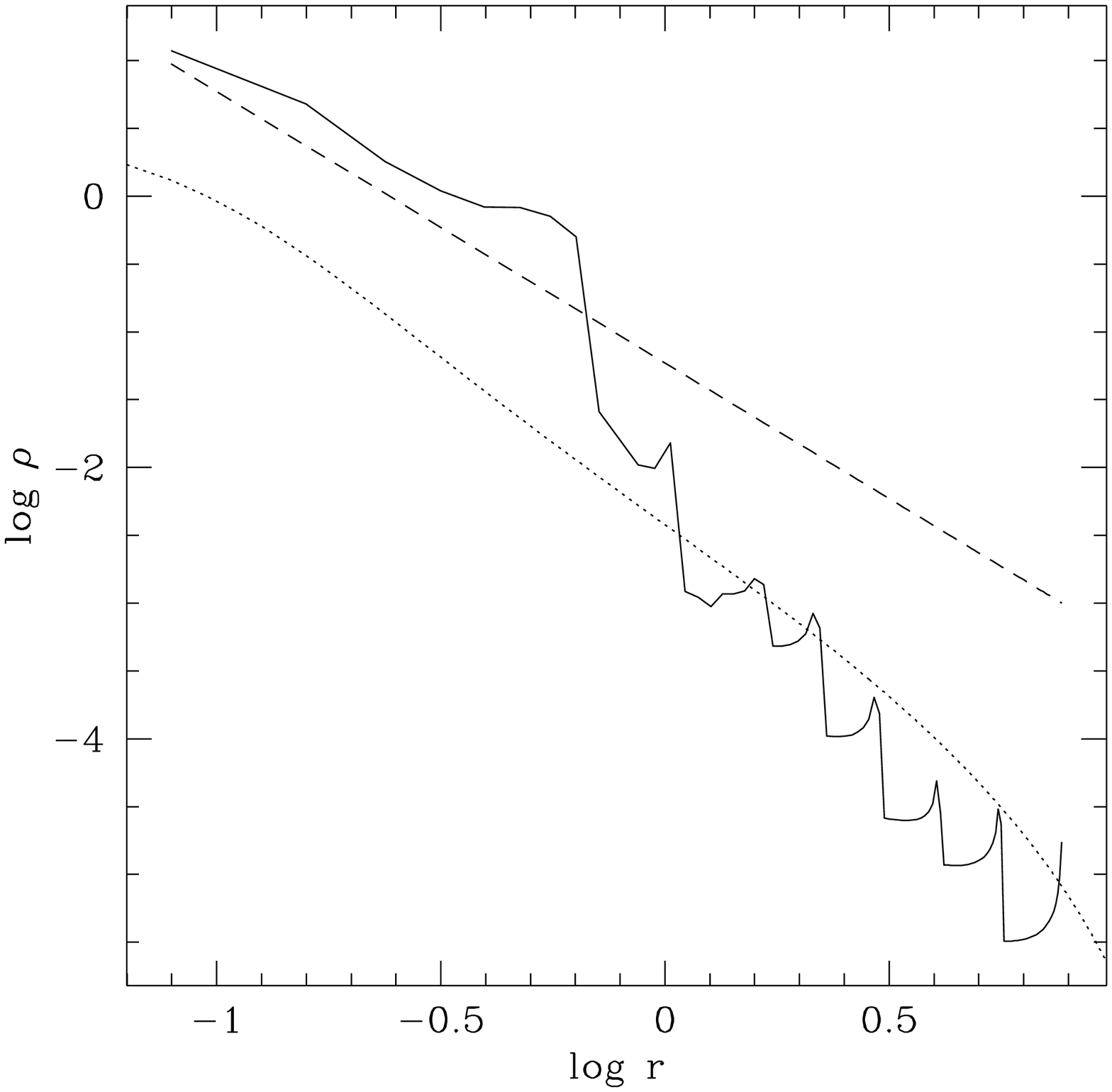}}
		\hbox{\epsfxsize=\hwidth\epsfbox{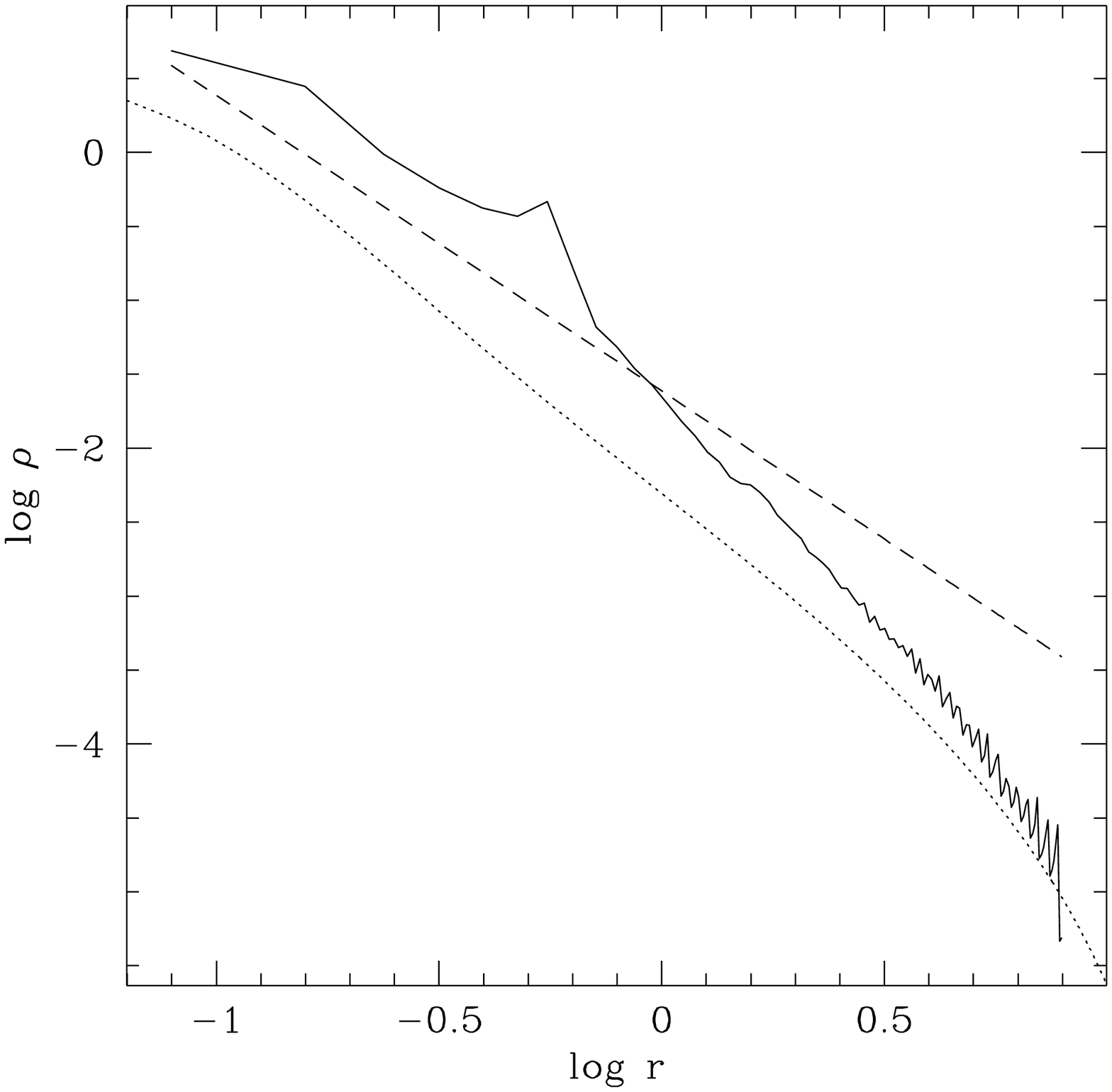}}
	}
	\hbox{
		\hbox{\epsfxsize=\hwidth\epsfbox{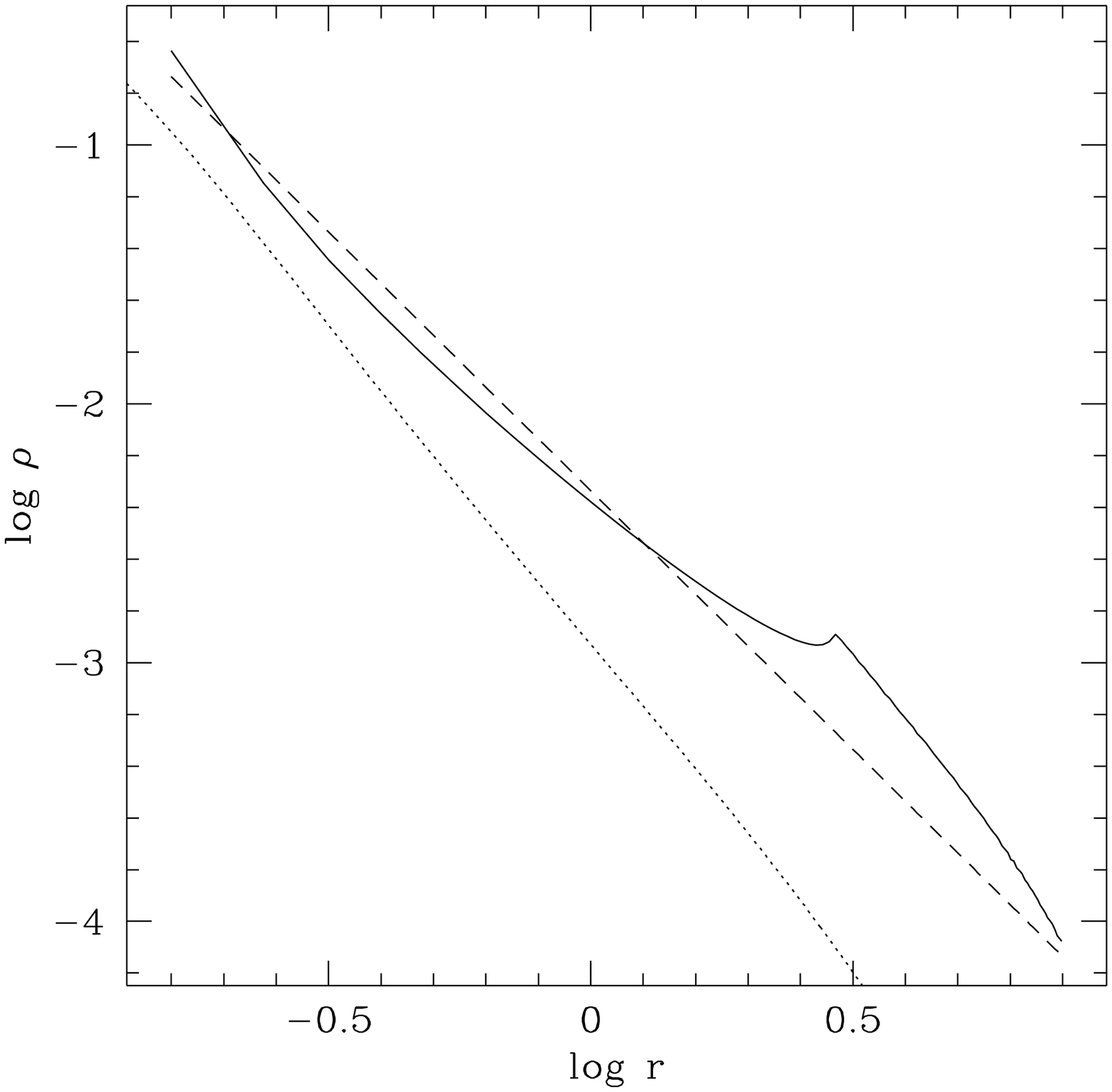}}
		\hbox{\epsfxsize=\hwidth\epsfbox{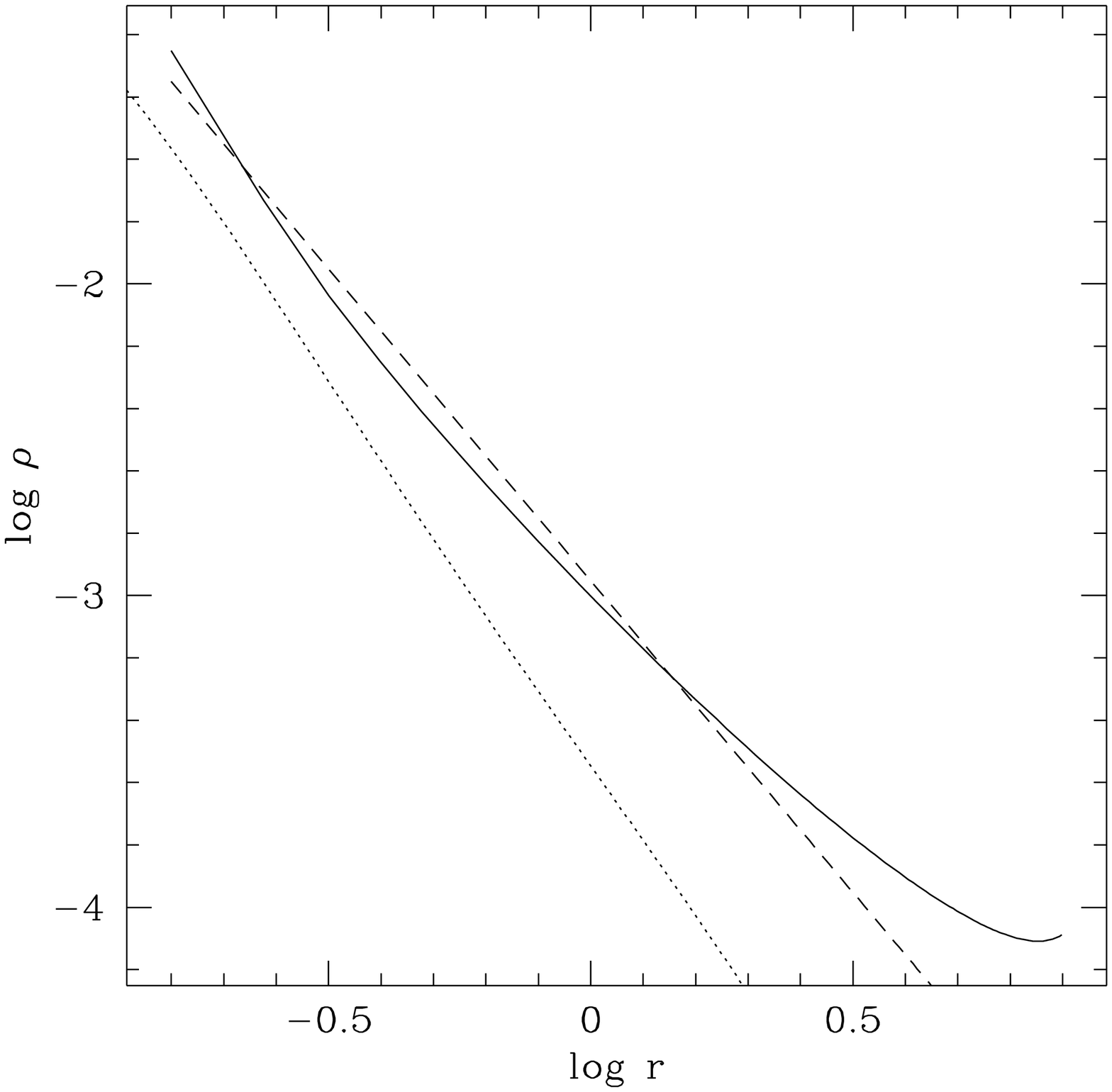}}
	}
}}
\caption{\label{fig:vmloss} Distribution of stripped material from
disrupting secondary (solid curve) compared with $r^{-2}$ profile
(dashed) and background density (dotted) at the four mass ratios
$10^{-1}, 10^{-2}, 10^{-3}, 10^{-4}$ (right to left, top to bottom).
These use the virial scaling and dimensionless units
(\S\protect{\ref{sec:scale}}).}
\end{center}
\end{figure}

As the secondary is stripped and disrupted, its stars preserve the
instantaneous orbit and build up the primary as suggested by Richstone
(1976\nocite{Rich:76}).  The relative density distributions for
secondaries on eccentric orbits, $\kappa=0.1$, and the four mass
ratios are shown in Figure \ref{fig:vmloss}.

The higher-mass secondaries, $M_{ratio}=10^{-1}$ and $10^{-2}$, lose
mass quickly.  Material is lost most quickly at pericenter and
individual episodes of mass loss during each orbit are visible in the
outer galaxy.  For $M_{ratio}=10^{-1}$, 90\% of the mass is lost
within the half-mass radius of the primary and 20\% within 4 core
radii.  Overall, the remnant profile is steeper than the primary and
could be a significant contributor to the inner light after a few such
events.  The two low mass ratio cases, $M_{ratio}=10^{-3}$ and
$10^{-4}$, lose mass more gradually and the distribution of stars lost
in the outer primary is more extended than the primary, approximating
a $r^{-2}$ distribution.  Both have guiding center radii larger the
primary half-mass radius and have lost roughly 80\% and 50\% of their
total mass at the point depicted.

Overall, these results suggest that {\it mass evaporated from the
secondary is distributed similarly to and maybe steeper than the
profile of the primary.}\/ After many accretion events, the merged
profile will be slightly more concentrated
(c.f. Fig. \ref{fig:vmloss}).

\subsection{Comparision with previous n-body simulation}

Unfortunately, there has been very little reported n-body work on
merging for self-consistent primary and secondary galaxies with masses
differing by orders of magnitude.  A search of the literature (Quinlan
1996/nocite{Quin:96}) revealed one similar n-body study by Balcells \&
Quinn (1990) who performed simulations with $M_{ratio}=0.1, 0.2$ for
rotating systems to explore the formation of counterrotating cores.
Although the their initial conditions and goals are sufficiently
different to prevent a precise comparison, the results here yield a
similar scenario.  The orbital decay is rapid with the largest
fraction the secondary evaporating in or near the core
(c.f. Fig. \ref{fig:mass1} and the first panel of
Figs. \ref{fig:orbdecay} and \ref{fig:vmloss}.).  Although Balcells \&
Quinn conclude that the dense core of the smaller galaxy {\it
survives} the orbital decay and settles in the core, I believe that we
describe the same the end result: a single merged profile rather than
a multiple nucleus system.  Clearly, this is subject is ripe for
additional work.

\section{Summary and discussion} \label{sec:summary}

The major conclusions of this work are as follows:
\begin{itemize}

\item Time-dependent heating will evaporate secondaries with mass
ratios as small as 1\% on {\it any} initial orbit well within a
galaxian age.  The difference with the naive prediction that the
denser satellite galaxies will invariably survive orbital decay is due
to the breakdown of the one-dimensional adiabatic invariant in
three-dimensional stellar systems as described in MW1.

\item Secondaries with mass ratios as small as 0.1\% on eccentric
orbits are significantly evolved and nearly evaporated.  Because
capture by dynamical drag will preferentially produce
high-eccentricity companions, this predicts a lower limit: {\it
captured secondaries with mass smaller than 0.1\% of the primary will
not evaporate.}

\item Captured satellites with mass ratios smaller than roughly 1\%
will not decay to the center in a galaxian age.  This limit may be
conservative---even lower mass secondaries will decay to the center
and evaporate---given evidence gravitational lens results
(Miralda-Escud\'e 1995\nocite{Mira:95}) that the optically inferred
mass profile of cD galaxies joins smoothing on the that of the cluster
in general.

\item The profile of the mass lost as the satellite decays is similar
to but slightly more concentrated than that of the original primary.
This implies that that the concentration of the cluster giant will
gradually increase after many mergers.

\item Altogether, we have the result that satellites which can fall to
the center of a cluster giant by dynamical friction are evaporated by
tidal heating in the process.  Disruption occurs near the center of
the primary; material from both cores combine into a single entity.
This suggests that true multiple nuclei giants should be rare.  This
scenario does not address the possibility that a massive accretion
event will lead to a nuclear gas accretion and a burst of star
formation (Hernquist \& Mihos 1995\nocite{HeMi:95}) and perhaps form a
second nucleus in situ.

\end{itemize}

\acknowledgments I thank Sandy Faber, Gerry Quinlan, Chigurupati
Murali, Doug Richstone and Scott Tremaine for comments and
suggestions.  This work was supported in part by NASA grant NAGW-2224,
NAG 5-2873 and the Sloan Foundation.

\appendix

\section{Orbit shocking} \label{sec:orbshock}

Shocking caused by an oscillatory perturbation is a straightforward
variant and is somewhat easier to compute than a one-shot adiabatic
disturbance described in W2.  For example, if a cluster is
dynamically part of the thick disk, then the perturbation has the form
$V_p=g(t)z^2$ where then $g(t)$ is a periodic function of time.  The
function $g(t)$ may be expanded as a Fourier series in its vertical
oscillation period, $P$:
\begin{equation} \label{eq:sum}
	g(t) = \sum_{k=-\infty}^\infty g_k e^{ik\omega t},
\end{equation}
where $\omega=2\pi/P$ and
\begin{equation}
	g_k = {1\over P}\int^P_0 dt\,g(t) e^{-ik\omega t}.
\end{equation}
The Laplace transform of this Fourier series is trivial and is:
\begin{equation}
	{\hat g} = \sum_{k=-\infty}^\infty {g_k\over s-ik\omega}.
\end{equation}
For physical scenarios (e.g. smooth and continuous mass profiles),
$g_k$ will converge rapidly with increasing $|k|$.  The calculation is
analogous for orbit shocking with the following changes:
\begin{enumerate}
\item The potential perturbation expansion will be more general than
the $\propto z^2$ dependence and include all 2nd order moments (all
$Y_{2m}$ terms);
\item The Fourier expansion of $g(t)$ will have two indices corresponding
to the radial and azimuthal periods of the cluster orbit.
\end{enumerate}
See MW2 for details.

We are only interested in the long-term secular change in the
distribution function, after any transients have decayed.  Following
W2, we Fourier-Laplace transform the perturbed Boltzmann equation.
The secular contribution is second-order distribution function and the
inverse Laplace transform leads to the desired result.  The details of
the function $g(t)$ are not important which allows one to eliminate it
altogether (MW2).  Alternatively, one can choose a convenient form for
$g(t)$, such as a square pulse, and perform the transforms explicitly.
Either way for time scales large compared to the stellar orbital
times, the secular change due to heating becomes
\begin{displaymath}
	{d\,f_2\over dt} = \pi \sum_{k,{\bf l}} |g_k|^2 \,{\bf
	l}\cdot{\partial\over\partial{\bf I}} \left(V_{t\,\bf
	l}V_{t\,-\bf l}\right) \,{\bf l}\cdot{\partial
	f_o\over\partial{\bf I}} \delta\left(k\omega+{\bf
	l\cdot\Omega}\right)
\end{displaymath}
where $V_{t\,\bf l}$ denotes the action-angle transform of the tidal
potential.  As described in MW2, this expression has the form:
\begin{equation}
	{\partial f\over\partial t} = {d\,f_2\over dt} = {\partial\over\partial
			E}\left\{A(E){\partial f\over\partial
			E}\right\}
\end{equation}
which may be solved by standard techniques (e.g. Crank-Nicholson or
Chang-Cooper 1970\nocite{ChCo:70} schemes).

MW3 will describe the effects disk-shocking, orbit shocking and
thick-disk shocking on the galactic population of globular clusters.
		
\section{Tidal perturbation}

To compute the effect of an orbit in a galaxy, the Galactic potential
may be expanded in the cluster frame.  The force (inertial) is:
\begin{eqnarray}
	{\bf F}_{t} &=& -\left.\nabla\Phi\right|_{{\bf R} + {\bf r}} +
\nabla\Phi|_{\bf R} \\
	F_i &\approx& -\left.\sum_j{\partial^2\Phi\over\partial x_i\partial
x_j}\right|_{R=R(t)} x_j,
\end{eqnarray}
where ${\bf R}$ describes the cluster and ${\bf r}$ the position of a
star relative to the cluster.  The tidal potential then follows
directly:
\begin{equation} \label{eq:tidal}
	V_{t} = {1\over2}\left.\left\{
	\left({d^2\Phi\over dR^2} - {1\over R}{d\Phi\over dR}\right)
	{  ({\bf R}\cdot{\bf x})^2\over R^2 } + 
	{1\over R}{d\Phi\over dR}\ r^2 \right\}\right|_{\bf R=R(t)} .
\end{equation}
Expanding equation (\ref{eq:tidal}) in spherical harmonics, perturbed
quantities may be computed as outlined in \S\ref{sec:orbshock}.  The
non-inertial velocity-dependent forces are not easily incorporated
into a potential and have been ignored here.

\end{document}